\documentclass[reprint,longbibliography,preprintnumbers,nofootinbib,amsmath,amssymb,aps,onecolumn]{revtex4-2}
\usepackage{dblfloatfix}
\usepackage{graphicx}
\usepackage{dcolumn}
\usepackage{bm}
\usepackage{color}
\usepackage{amssymb}
\usepackage{pifont}
\definecolor{green2}{rgb}{0.0, 0.5, 0.0}
\usepackage[dvipsnames]{xcolor}
\usepackage[colorlinks = true,
            linkcolor = blue,
            urlcolor  = blue,
            citecolor = green,
            anchorcolor = blue]{hyperref}
\usepackage{verbatim}
\makeatletter
\usepackage{multirow, graphicx,amssymb,url,mathrsfs,amsmath}
\usepackage{amsxtra,amstext,latexsym,dsfont,amsfonts}
\usepackage{color,eucal}
\usepackage[dvipsnames]{xcolor}
\usepackage{float}
\usepackage{colortbl}
\usepackage{multirow}
\usepackage{tikz}
\usetikzlibrary{tikzmark}
\usetikzlibrary{arrows}

\newcommand{\Rmnum}[1]{\expandafter\@slowromancap\romannumeral #1@}
\usepackage{tabularx, array}
\newcolumntype{L}[1]{>{\raggedright\arraybackslash}p{#1}}
\newcolumntype{C}[1]{>{\centering\arraybackslash}p{#1}}
\newcolumntype{R}[1]{>{\raggedleft\arraybackslash}p{#1}}
\makeatother

\makeatletter
\newcommand{\myBig}{\bBigg@{1.75}}
\makeatother

\begin{document}

\title{\Large The local thermodynamic instability from negative susceptibility in a holographic superfluid with nonlinear terms}

\author{Yu-Xiang Cao}
\affiliation{Center for Gravitation and Astrophysics, Kunming University of Science and Technology, Kunming 650500, China}
\author{Hui Zeng}
 \email{zenghui@kust.edu.cn}
 \email{Corresponding author.}
\affiliation{Center for Gravitation and Astrophysics, Kunming University of Science and Technology, Kunming 650500, China}
\author{Zhang-Yu Nie}
 \email{niezy@kust.edu.cn}
 \email{Corresponding author.}
\affiliation{Center for Gravitation and Astrophysics, Kunming University of Science and Technology, Kunming 650500, China}

\begin{abstract}
The local thermodynamic stability from the charge susceptibility of a holographic superfluid model at finite superfluid velocity is studied in the probe limit. 
Previous studies show that beyond a finite value of the superfluid velocity, the superfluid phase transition in the grand canonical ensemble becomes first order. We further reveal that in the canonical ensemble, the superfluid phase transition is still second order, and the difference indicates a section with negative susceptibility which means local thermodynamic instability beyond this superfluid velocity. However, we also meet the ``cave of wind'' behavior at larger superfluid velocity which complicates the phase diagram. We further study the influence of the two nonlinear terms $\lambda|\psi|^4$ and $\tau|\psi|^6$ with parameters $\lambda$ and $\tau$ on the condensate curves, and set appropriate values of $\lambda$ and $\tau$ to remove the "cave of wind" region in the canonical ensemble to get a more elegant phase diagram and better represent the region with such instability, which is possible to be used to realize spontaneous formation of vortexes and quantum turbulence.
\end{abstract}
\maketitle
\tableofcontents

\section{Introduction}\label{Introduction}
As a strong-weak duality, the gauge-gravity duality provides us an effective instrument to understand the nature of strongly coupled gauge field theory by studying its weakly coupled gravitational dual \cite{Maldacena:1997re,Gubser:1998bc,Witten:1998qj}. In the last decade, this holographic duality has been widely applied to condensed matter systems. Considering the (3 + 1)-dimensional Schwarzschild anti-de Sitter (AdS) black hole coupled with a Maxwell-complex scalar field, the holographic s-wave superconductor model was first built by Hartnoll, Herzog and Horowitz to reproduce properties of a (2 + 1)-dimensional superconductor~\cite{Hartnoll:2008vx}. With more general charged fields in the bulk, the p-wave and d-wave superconductors as well as competition and coexistence between the various orders are also realized in holography~\cite{Gubser:2008wv,Cai:2013pda,Cai:2013aca,
Chen:2010mk,Benini:2010pr,Basu:2010fa,Musso:2013ija,Nie:2013sda,Donos:2013woa,Li:2017wbi,Nie:2015zia,Nie:2014qma,Amado:2013lia}.

The original s-wave model is also developed to study states with nonzero superfluid velocity~\cite{Basu:2008st,Herzog:2008he}, where it is found that the superfluid phase transition becomes 1st order at large value of superfluid velocity in the grand canonical ensemble. The authors of Ref.~\cite{Arean:2010zw} explored more rich superfluid phase structure with different bulk spacetime dimensions and various values of the scalar mass parameter. Similar results also arise in the case of excited states~\cite{Wang:2023blh}, as well as in the holographic p-wave~\cite{Wu:2014bba,Wu:2014cfa,Zeng:2010fs} superfluid models. The holographic superfluid solutions with nonzero superfluid velocity are also observed in various studies on the time dependent evolution of non equilibrium states~\cite{Adams:2012pj,Sonner:2014tca,Chesler:2014gya,Zeng:2019yhi,Xia:2021xap,Xia:2024ton,Yang:2024hom}, where the superfluid velocity is inhomogeneous in most cases. Among these studies, the Kibble-Zurek Mechanism (KZM) is verified in holography in Refs.~\cite{Sonner:2014tca,Chesler:2014gya,Zeng:2019yhi}. However, the break-down of the Kibble-Zurek scaling is also observed for sufficiently fast quenches~\cite{Sonner:2014tca,Chesler:2014gya}, and new universality in case of very fast quenches is discovered from holography in Ref.~\cite{Xia:2021xap} and later investigated systematically in Ref.~\cite{Zeng:2022hut}. 

Sometimes a uniform initial state is deformed to be inhomogeneous with formation and evolution of domain walls~\cite{Janik:2017ykj,Attems:2019yqn,Attems:2020qkg,Zhao:2023ffs,Chen:2024pyy}. In a recent study on the Quasi-Normal Modes of the holographic superfluid solutions, it is claimed that when the condensate curve grows to the different direction in the grand canonical ensemble with the condensate curve in the canonical ensemble, the superfluid phase becomes unstable under inhomogeneous perturbations~\cite{Zhao:2022jvs}. This kind of inhomogeneous instability is soon confirmed by the successive study on the full dynamical evolution in the same model, where the formation and evolution of bubble configurations are presented~\cite{Zhao:2023ffs}. Typical cases of such a situation are usually accompanied by 1st order phase transitions, which indicates studying the linear instability problem in the holographic superfluid with a large superfluid velocity where the phase transition becomes 1st order.

Therefore in this work, we study the linear stability of the s-wave holographic superfluid model at finite superfluid velocity. As a beginning step, we focus on the linear thermodynamic instability related to the negative value of charge susceptibility $\partial \rho/\partial \mu$, which is obvious from comparing the condensate curves in the grand canonical ensemble and the canonical ensemble based on the analysis in~\cite{Zhao:2022jvs}. We also extend the study on the effects of fourth and sixth power nonlinear terms to finite values of the superfluid velocity, which makes it easy to improve the final $S_x-\rho$ phase diagram to focus on the instability region that is useful in future study on possible realization of quantum turbulence or superfluid vortexes and patterns.
This work is organized as follows. In Section 2, we will construct the holographic s-wave superfluid model with nonlinear self-interaction terms in the AdS black hole background. In Section 3, we will investigate the condensates of the scalar field and the phase transition in the holographic superfluid both for the grand canonical ensemble and canonical ensemble. We will conclude in the last section 4 with our main results and present a short outlook as well as some perspectives for the future.
\section{Holographic setup}  \label{sec1}
We consider a holographic s-wave superfluid model with nonlinear self-interaction terms in (3+1) dimensional asymptotic AdS spacetime~\cite{Zhao:2022jvs,Zhao:2023ffs,Zhao:2024jhs}, which is a simple extension of the HHH model \cite{Hartnoll:2008vx}. The bulk action is given by
\begin{align}
	S=&\,S_{M}+S_{G}~,\label{Lagall}\\
	S_G=&\,\frac{1}{2\kappa_g ^2}\int d^{4}x\sqrt{-g}(R-2\Lambda)~,\label{Lagg}\\
	S_M=&\,\int d^{4}x\sqrt{-g}\Big(-\frac{1}{4}F_{\mu\nu}F^{\mu\nu}-D_{\mu}\Psi^{\ast}D^{\mu}\Psi-m^{2}\Psi^{\ast}\Psi-\lambda(\Psi^{\ast}\Psi)^{2}-\tau(\Psi^{\ast}\Psi)^{3}\Big)~.\label{Lagm}
\end{align}
Here $F_{\mu\nu}=\nabla_{\mu}A_{\nu}-\nabla_{\nu}A_{\mu}$ is the strength tensor of the U(1) gauge field $A_{\mu}$, $D_{\mu}\Psi=\nabla_{\mu}\Psi-i q A_\mu\Psi$ is the standard covariant derivative term of the complex scalar field $\Psi$ with charge $q$ and mass $m$. $\Lambda=-3/L^2$ is the negative cosmological constant and $L$ is the AdS radius. The additional two terms appearing in the matter Lagrangian, Eq.\eqref{Lagm}, $-\lambda(\Psi^{\ast}\Psi)^{2}$ and $-\tau(\Psi^{\ast}\Psi)^{3}$, introduce nonlinear self-interactions for the bulk scalar field $\psi$. Based on these two additional terms, the phase structure
of the model becomes more abundant.

Since we focus on the possible instability from negative charge susceptibility, it is sufficient to work in the probe limit, where the back ground metric is fixed to be the (3 + 1)-dimensional planar Schwarzschild-AdS black hole in the form
\begin{align}
	&\ ds^{2}=-f(r)dt^{2}+\frac{1}{f(r)}dr^{2}+\frac{r^{2}}{L^2}\left(dx^{2}+dy^{2}\right)~,\label{metric1}\\
	&\ f(r)=\frac{r^{2}}{L^{2}}-\frac{M r_h^{3}}{r}~,\label{metric2}
\end{align}
where $r=r_{h}$ is the black hole horizon. The Hawking temperature of this black-brane solution is given by
\begin{align}
    T= \frac{3 r_h}{4\pi L^{2}}~.
\end{align}

In order to construct the holographic superfluid solutions, we set the ansatz of the matter fields as
\begin{align}\label{ansatz}
    \Psi=\psi(r)~, \quad A_{\mu}=\left(\phi(r),0,A_{x}(r),...\right)~.
\end{align}
The equations of motion are
\begin{align}
 	\psi''+\left(\dfrac{2}{r}+\dfrac{f'}{f}\right)\psi'+\left(\dfrac{q^2 \phi^2}{f^2}-\dfrac{q^{2}A_{x}^2}{r^2 f}-\dfrac{m^2}{f}\right)\psi-\dfrac{2\lambda}{f}\psi^3-\dfrac{3\tau}{f}\psi^5=0~,\label{eqpsi}\\
 	\phi''+\dfrac{2}{r}\phi'-\dfrac{2q^{2}\psi^{2}}{f}\phi=0~,\label{eqphi}\\
 	A_x''+\dfrac{f'}{f}A_x'-\dfrac{2q^{2}\psi^2}{f}A_x=0~,\label{eqA_x}
\end{align}
where the prime represents the derivative of r.

In the remainder of this paper, we seek solutions with negative charge susceptibility, which can be determined by comparing the condensate curves in the grand canonical ensemble with those in the canonical ensemble. For simplicity and without lose of generality, we set $m^2=-2$, $q=1$, $r_{h}=1$ and $L=1$. 

To solve the equations of motion numerically, we need to specify the boundary conditions. Near the horizon $r=r_{h}$, the fields $\psi$ and $A_{x}$ are regular, but $\phi$ satisfies $\phi(r_{h}) = 0$. Therefore the expansions near the horizon are
\begin{align}
    &\phi(r)=\phi_{1}(r-r_{h})+\mathcal{O}((r-r_{h})^{2})~,\\
    &A_x(r)=A_{x0}+A_{x1}(r-r_{h})+\mathcal{O}(r-r_{h})~,\\
    &\psi(r)=\psi_{0}+\psi_{1}(r-r_{h})+\mathcal{O}(r-r_{h})~.
\end{align}
The expansions near the AdS boundary $r \rightarrow{\infty}$ are
\begin{align}
	&\phi(r)=\mu-\frac{\rho}{r}+...~,\qquad \\
	&A_x(r)=S_x-\frac{J_x}{r}+...~,\qquad \\
	&\psi(r)=\frac{\psi^{(1)}}{r}+\frac{\psi^{(2)}}{r^2}+...~.
\end{align}
According to the AdS/CFT dictionary. $\mu$ and $\rho$ are interpreted as the chemical potential and charge density, while $S_{x}$ and $J_{x}$ are identified as the superfluid velocity and current in the dual field theory, respectively.  As in Ref.\cite{Herzog:2008he}, we also define the Lorentz invariant quantity $\mu_{L}=\sqrt{\mu^2-S_{x}^2}$, $\rho_{L}=\sqrt{\rho^2-J_{x}^2}$.
With our choice of the value of mass parameter, both $\psi^{(1)}$ and $\psi^{(2)}$ are normalizable and can be used to define the dual scalar operator. We choose the standard quantization which means that $\psi^{(1)}$ is regarded as the source term for the boundary operator while $\psi^{(2)}=\big\langle\mathcal{O}\big\rangle$ is regarded as the vacuum expectation value. We set $\psi^{(1)}=0$ as the source free boundary condition to obtain the solutions corresponding to the spontaneous symmetry breaking of the boundary global U(1) symmetry.

In order to study the global thermodynamic stability, we should calculate the thermodynamic potential such as the grand potential $\Omega$ of the boundary states. In the probe limit, the contribution from the metric part is always fixed, and we express the matter contribution to the Euclidean action $S_{ME}$ as
\begin{align}
	S_{ME}&=\int d^{3}x(\frac{1}{2}r^2\phi\phi^{'}-\frac{1}{2}fA_{x}A_{x}'-r^2f\psi\psi^{'})\big|_{r=\infty}+\int d^{4}x(-\frac{q^{2}r^{2}\phi^{2}\psi^{2}}{f}+q^{2}\psi^{2}A_{x}^{2}+r^2\lambda\psi^4+2r^{2}\tau\psi^{6})\\
	&=-\frac{V_2}{T}\Big[\frac{1}{2}\mu\rho-\frac{1}{2}S_{x}J_{x}+\int(-\frac{q^{2}r^{2}\phi^{2}\psi^{2}}{f}+q^{2}\psi^{2}A_{x}^{2}+r^2\lambda\psi^4+2r^{2}\tau\psi^{6})dr\Big]~.
\end{align}
With the integration over the Euclidean time and the boundary spacial coordinates $\int d^{3}x=V_2/T$, the grand potential $\Omega_s$ of the superfluid phase is given by
\begin{align}
	\frac{\Omega_s}{V_2}=\frac{TS_{ME}}{V_2}=-\frac{\mu\rho}{2}+\frac{S_x J_x}{2}-\int_{r_h}^{\infty}(-\frac{q^{2}r^{2}\phi^{2}\psi^{2}}{f}+q^{2}\psi^{2}A_{x}^{2}+r^2\lambda\psi^4+2r^{2}\tau\psi^{6})dr~,
\end{align}
where, $V_{2}$ is the spatial volume of the boundary manifold.
\section{The condensate curves and the phase diagrams}\label{sec2}
\subsection{The condensate curves and the phase diagram without nonlinear self-interaction terms}\label{subsection1}
\begin{figure}
\includegraphics[width=0.3\columnwidth]{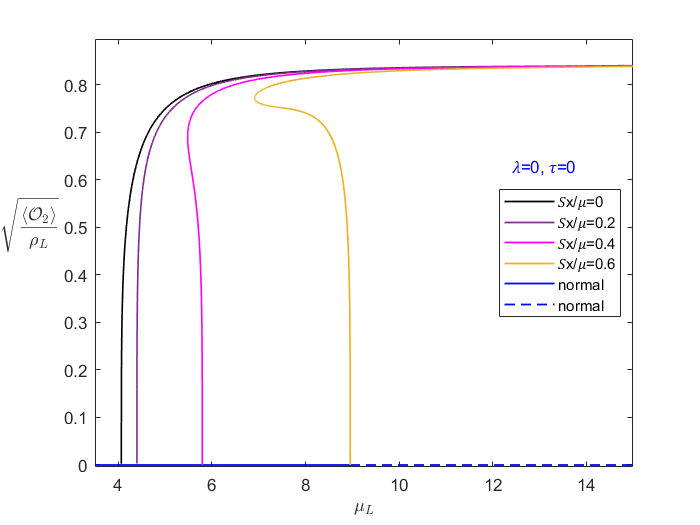}
\includegraphics[width=0.3\columnwidth]{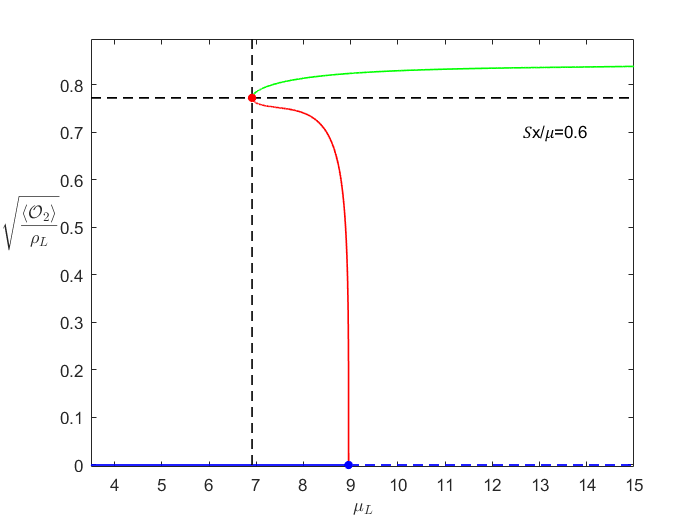}
\includegraphics[width=0.3\columnwidth]{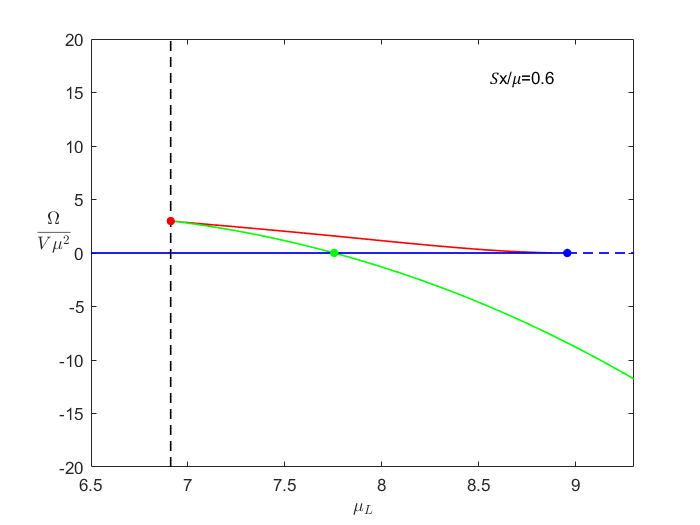}
\caption{The condensate and grand potential for solutions with $\lambda=0$ and $\tau=0$ in the grand canonical ensemble.
\textbf{Left: }the condensate as a function of the chemical potential for different values of $S_{x}/\mu$, i.e., $S_{x}/\mu$=0, 0.2, 0.4 and 0.6.
\textbf{Middle: }the condensate with the superfluid velocity $S_{x}=0.6\mu$. 
\textbf{Right: }The grand potential with the superfluid velocity $S_{x}=0.6\mu$.
 The blue point indicates the position of the quasi critical point, the red point indicates the turning point and the green point indicates the phase transition point of the first order phase transition.}\label{1_OG_Mu}
\end{figure}

Without the nonlinear terms ($\lambda=\tau=0$), it is found in Ref.~\cite{Herzog:2008he} that in the grand canonical ensemble, the superfluid phase transition becomes 1st order when the superfluid velocity is larger than the special value $S_{x}=0.27\mu$. We first repeat these results and show them concretely in Figure~\ref{1_OG_Mu}. The left panel of Figure~\ref{1_OG_Mu} shows the condensate curves with various values of the superfluid velocity $S_x/\mu$. It is clear that when the superfluid velocity increases, the critical point have a higher value of chemical potential which means the superfluid phase transition becomes harder. Further more, the superfluid phase transition is 2nd order when the superfluid velocity is below the special value $S_{x}=0.27\mu$, and becomes 1st order when $S_{x}>0.27\mu$. We further take the case of $S_{x}=0.6\mu$ as an example and plot its condensate curve as well as the grand potential density in the middle and right panels of Figure~\ref{1_OG_Mu}. We can see from the middle panel that the condensate grows to the left from the ``quasi critical point''~\cite{Zhao:2023ffs,Zhao:2024jhs} marked by the blue dot and then grows to the right after the turning point marked by the red dot. The grand potential curve in the right panel shows a standard swallow tail shape of the first order phase transitions and the phase transition point of the 1st order phase transition is the intersection point of the green and blue grand potential curves marked by the green dot. The grand potential as well as the landscape analysis~\cite{Zhao:2022jvs} show that in the grand canonical ensemble, the red section of the solutions are unstable while the green section of superfluid solutions are stable.
\begin{figure}
\includegraphics[width=0.45\columnwidth]{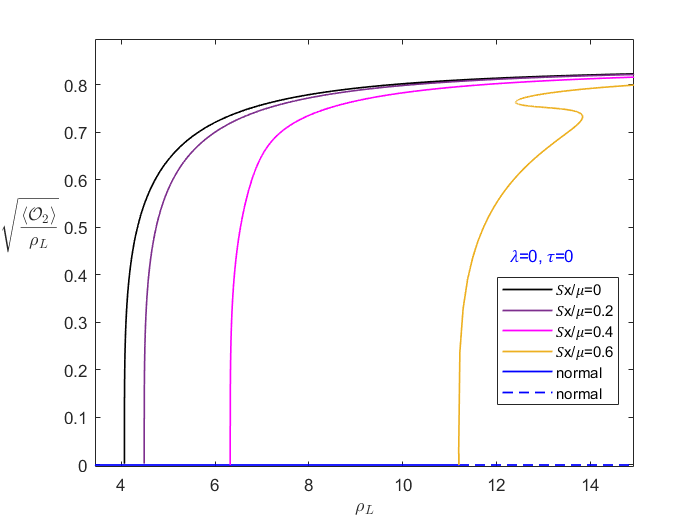}
\includegraphics[width=0.45\columnwidth]{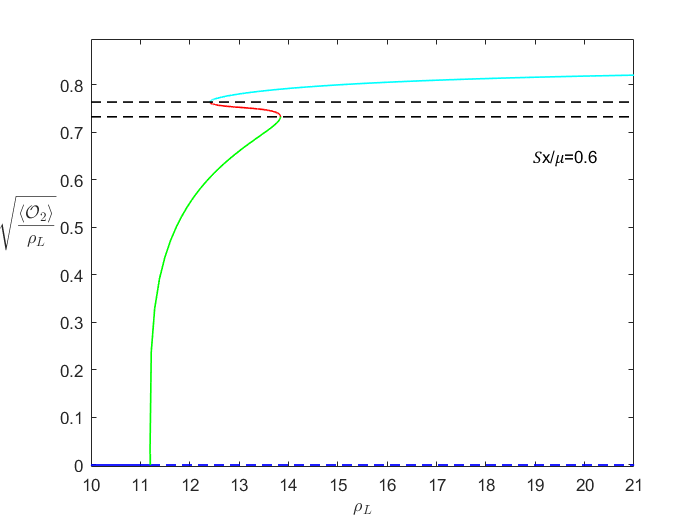}
\caption{The condensate for solutions with $\lambda=0$ and $\tau=0$ in the canonical ensemble.
\textbf{Left: }the condensate as a function of the charge density for different values of $S_{x}/\mu$, i.e., $S_{x}/\mu$=0, 0.2, 0.4 and 0.6.
\textbf{Right: }the condensate with the superfluid velocity $S_{x}=0.6\mu$.
In the right panel, the solutions on the solid green line and solid sky blue line are stable, while the solutions on the solid red line are unstable. }\label{1_O_Rho}
\end{figure}

In order to study the linear instability from the thermodynamic analysis, we also present the condensate curves for the same solutions in the canonical ensemble in Figure~\ref{1_O_Rho}. In the left panel of Figure~\ref{1_O_Rho}, we plot the condensate curves for different values of superfluid velocity, where we see that the phase transition from the normal phase to the superfluid phase is always 2nd order. However, when the superfluid velocity is larger than another special value $0.51\mu$, the condensate curve show the ``cave of wind'' behavior, which indicates a first order phase transition between the superfluid phase with larger condensates and the superfluid phase with smaller condensates. We take the case of $S_x=0.6\mu$ as an example and show the condensate more clearly in the right panel of Figure~\ref{1_O_Rho}, where the red section of superfluid solution is unstable.

Because in the region $S_x>0.27\mu$, the superfluid phase transition from the normal phase becomes 1st order in the grand canonical ensemble, but still keep 2nd order in the canonical ensemble, the charge susceptibility $\partial \rho/\partial \mu$ becomes negative for the section with small condensates. From the study in Refs.~\cite{Zhao:2022jvs,Zhao:2023ffs}, the region with negative susceptibility will suffer linear instability of inhomogeneous perturbations at large scale. Therefore we compare the condensate curves in the two ensembles to locate the region with negative susceptibility in the phase diagram in the canonical ensemble.

\begin{figure}
\centering
\includegraphics[width=0.45\columnwidth]{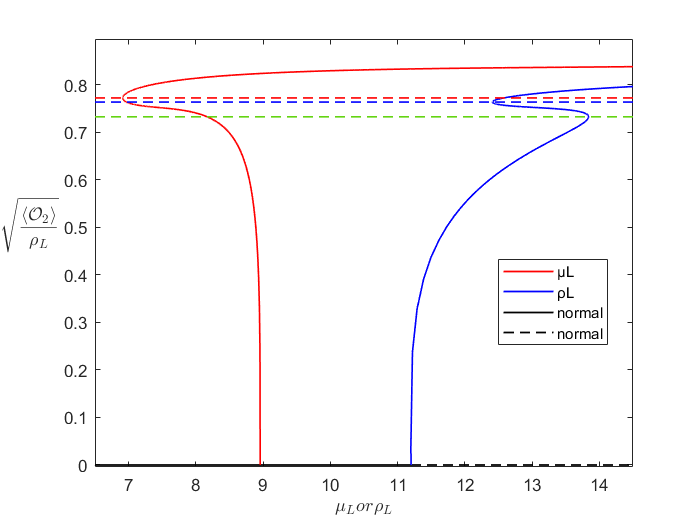}
\caption{The condensate curves in the grand canonical ensemble (solid red line) and the canonical ensemble (solid blue line) with $\lambda=0$, $\tau=0$ for superfluid velocity $S_{x}=0.60\mu$.
The  horizontal dashed red line indicates the place of the turning point of the solid red curve, while the horizontal dashed green and dashed blue lines indicate the two turning points of the solid blue curve.}
\label{1_MuRho_0.6}
\end{figure}

In the region $\S_x<0.27\mu$, both the phase transitions in the grand canonical ensemble and in the canonical ensemble are 2nd order, and the charge susceptibility is always positive. When $0.27\mu<\S_x<0.51\mu$, the region between the critical point and the point corresponding to the turning point in the grand canonical ensemble get negative value of the charge susceptibility and suffers from the inhomogeneous instability in the canonical ensemble. The case with $S_x>0.51\mu$ is more complicated, because there is an additional 1st order phase transition between the superfluid solution with small condensate and the superfluid solution with larger condensate. To give a more concrete description for this case, we plot the condensate curve with respect to the chemical potential $\mu_L$ (in the grand canonical ensemble) and the condensate curve with respect to the charge density $\rho_L$ (in the canonical ensemble), respectively in Figure~\ref{1_MuRho_0.6}.

In Figure~\ref{1_MuRho_0.6}, the solid red curve on the left side show the condensate in the grand canonical ensemble, while the solid blue curve on the right side show the COW behavior of the condensate in the canonical ensemble. We further use the horizontal dashed red line to mark the place of the turning point of the solid red curve, while use the dashed green and blue lines to mark the two turning points of the solid blue curve. Then we see clearly the charge susceptibility $\partial\rho_L/\partial\mu_L$ is negative below the dashed green line and between the dashed red and dashed blue lines, where the solid red curve and the solid blue curve grow in different directions. Between the dashed green and the dashed blue lines, the charge susceptibility is positive, however, the solutions in this region in unstable from the landscape point of view in both the grand canonical ensemble and the canonical ensemble, where even homogeneous perturbations are unstable~\cite{Zhao:2022jvs}.

\begin{figure}
\includegraphics[width=0.30\columnwidth]{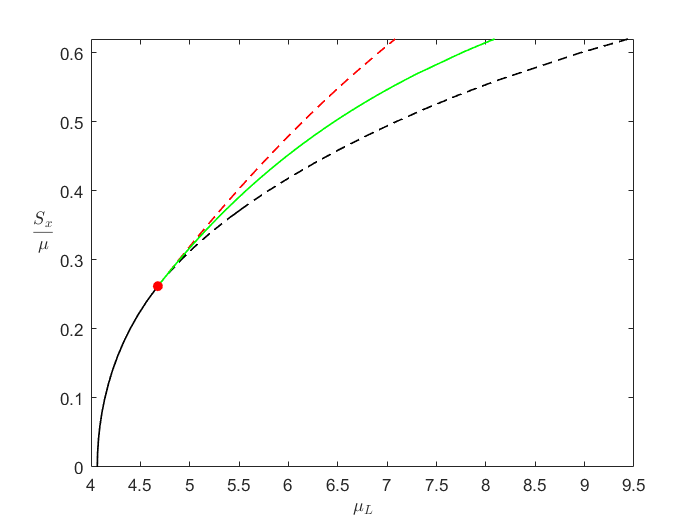} 
\includegraphics[width=0.30\columnwidth]{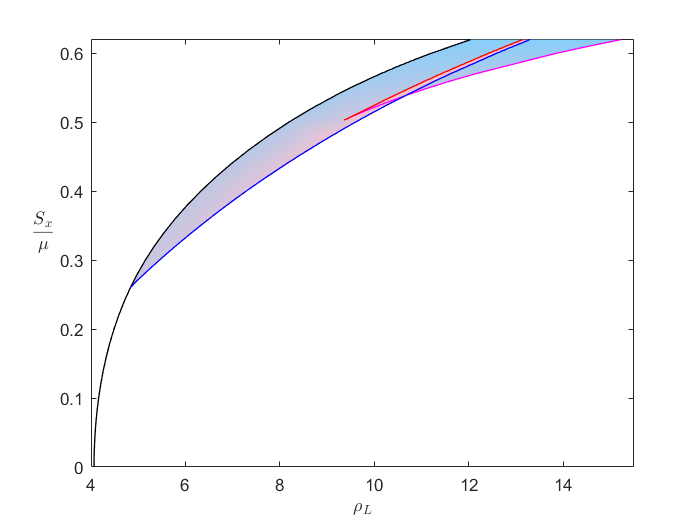}
\includegraphics[width=0.30\columnwidth]{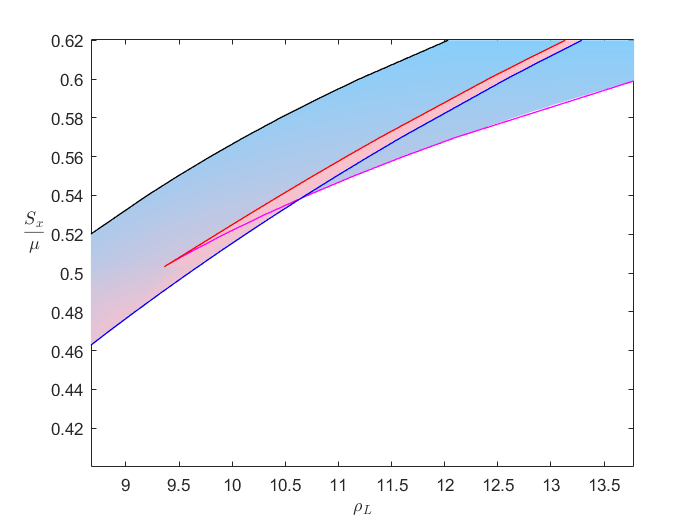}
\caption{The phase diagrams with $\lambda=0$ and $\tau=0$.
 \textbf{Left: }The phase diagram in the grand canonical ensemble. The solid black line indicates the critical points for the 2nd order superfluid phase transitions. The solid green line indicates the phase transition points for the first order superfluid phase transitions, while the dashed black and dashed red lines indicate the quasi-critical points and the turning points, respectively.
 \textbf{Middle: }The phase diagram in the canonical ensemble. 
 \textbf{Right: }A locally enlarged version of the phase diagram in the canonical ensemble. In the middle and right panels, the solid black line indicates the position of the critical points for the 2nd order superfluid phase transitions. The solid magenta line and solid red line indicate the two turning points of the COW condensate curves. The solid blue line is mapped from the dashed red line in the left panel which indicates the turning points of the condensate curves in the grand canonical ensemble. The colored region marked the domain with linear thermodynamic instabilities caused by the negative charge susceptibility.
 \label{1_ph_MuRho}
 }
\end{figure}

Based on the above analysis, we are able to locate the region with negative susceptibility in the $S_{x}/\mu-\rho_L$ phase diagram in the middle panel of Figure~\ref{1_ph_MuRho}. We also represent the $(\mu_L, S_{x}/\mu)$ phase diagram~\cite{Herzog:2008he} in the left panel of Figure~\ref{1_ph_MuRho} as a comparison. The right panel of Figure~\ref{1_ph_MuRho} is an enlarged version of the middle panel to show more details near the sharp tip. We can see from the left panel that the superfluid phase transition becomes first order when the superfluid velocity is larger than the vertical coordinate of the red dot, and the solid green line shows the position of the 1st order phase transition points while the dashed red line shows the turning points of the condensate curve in the grand canonical ensemble.

From the $S_x-\rho$ phase diagram presented by the middle and right panels of Figure~\ref{1_ph_MuRho}, it is obvious that the superfluid phase transition from the normal phase is always 2nd order. However, the complicity comes from the ``cave of wind'' behavior of the condensate curve, which indicates a 1st order phase transition between the superfluid phases with higher and lower condensates. Since we still meet some problems in calculating the free energy consistently, we only show the two turning points marked by the solid red line and the magenta line without giving the accurate position of the phase transition points. We further add the solid blue line, which is mapped from the dashed red line in the left panel and indicates the turning points in the grand canonical ensemble. Then the section of superfluid solutions between the critical points marked by the solid black line and the points on the solid blue line is believed to suffer from inhomogeneous linear instability in the canonical ensemble. When the superfluid velocity $S_x$ is larger than the vertical coordinate of the tip point on the red line, the situation becomes much more complicated because of the triply folded solutions in the region between the solid red line and the solid magenta line. From the previous analysis, the solid blue line in the middle and right panels of Figure~\ref{1_ph_MuRho} is mapped from solution on the dashed red line in Figure~\ref{1_MuRho_0.6}, which is always on the higher branch of the blue condensate curve in the canonical ensemble. In this case, both the lower branch of solutions below the dashed green line and the branch of solutions between the dashed blue and dashed red lines in Figure~\ref{1_MuRho_0.6} suffer from the instabilities related to the negative susceptibility. These two branches of solutions at different values of superfluid velocity are further mapped into the two regions, one region colored blue between the solid black line and the solid magenta line and the other region colored red between the solid red line and the solid blue line, in the $S_x-\rho$ phase diagram in the middle and right panels of Figure~\ref{1_ph_MuRho}. Both the blue and red regions and the region with gradient color show the domain with negative susceptibility concretely. We should notice that the red region is always overlap on the blue region at the same place, which make the resulting phase diagram rather complicated.
\subsection{The power of the nonlinear terms and the improved phase diagram}\label{subsection2}
In order to simplify the results, we resort to the powerful nonlinear terms with coefficients $\lambda$ and $\tau$ to remove the ``cave of wind'' behavior in the canonical ensemble, but keep the 1st order phase transition in the grand canonical ensemble. Then the resulting phase diagram in the canonical ensemble and the region with negative susceptibility become more elegant. We first study the influence of the two terms with coefficient $\lambda$ and $\tau$ on the condensate curves in both the grand canonical ensemble and the canonical ensemble, respectively.

\begin{figure}
\includegraphics[width=0.45\columnwidth]{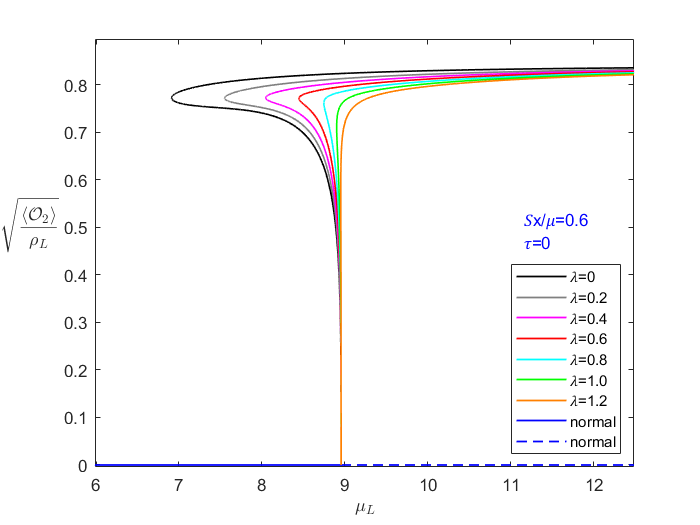} 
\includegraphics[width=0.45\columnwidth]{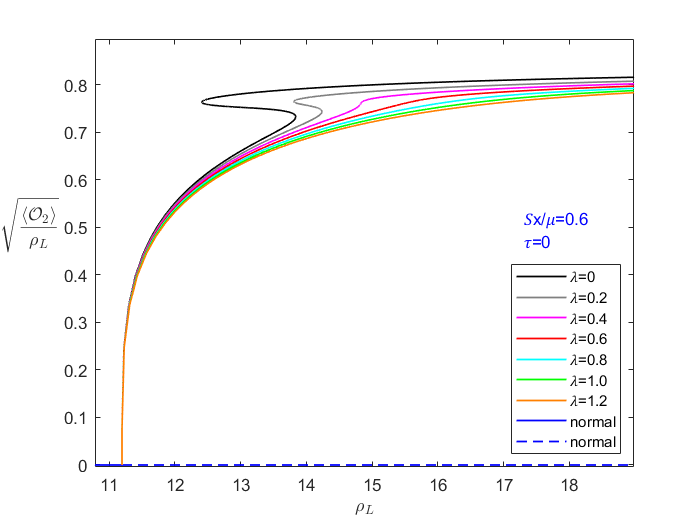}
\caption{The condensate curves in the grand canonical ensemble (\textbf{Left}) and in the canonical ensemble (\textbf{Right}) with $\tau=0$ and different values of $\lambda$.
}\label{2_lam_MuRho}
\end{figure}

In Figure~\ref{2_lam_MuRho}, we plot the condensate curves with various values of $\lambda$ and $S_x=0.6$ to show the influence of the fourth power term. The left panel shows the condensate curves in the grand canonical ensemble while the right panel shows the condensate curves in the canonical ensemble. We can see from these two plots that, with the increasing value of $\lambda$, the condensate curve changes from the black curve on the left to the colored curves on the right. The left panel shows that in the grand canonical ensemble, the 1st order phase transition with $\lambda=0$ is changed into a 2nd order phase transition at large values of $\lambda$, which shows a powerful control of the fourth power term on the order of phase transitions~\cite{Herzog:2010vz,Zhao:2022jvs,Zhao:2024jhs}. The right panel shows that the ``cave of wind'' behavior with $\lambda=0$ is also removed with a large value of $\lambda$, which shows that the fourth power term is also effective in tuning the 1st order phase transition between the superfluid phases with higher and lower condensates, as well as pushing the system into the supercritical region as in the case with zero superfluid velocity~\cite{Zhao:2024jhs}.

\begin{figure}
\includegraphics[width=0.45\columnwidth]{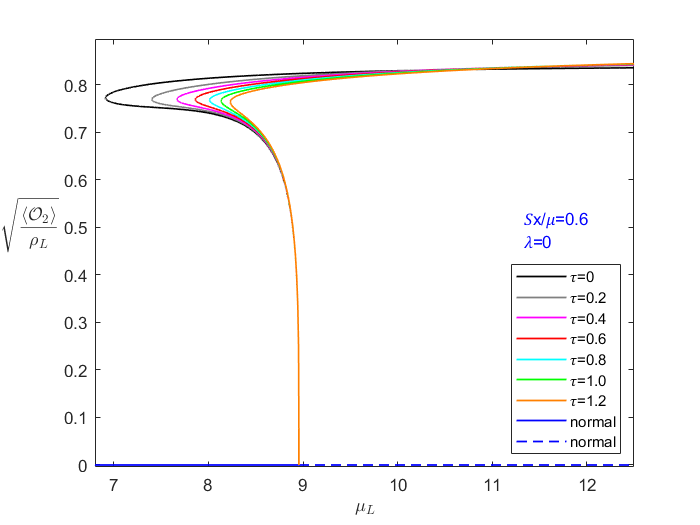} 
\includegraphics[width=0.45\columnwidth]{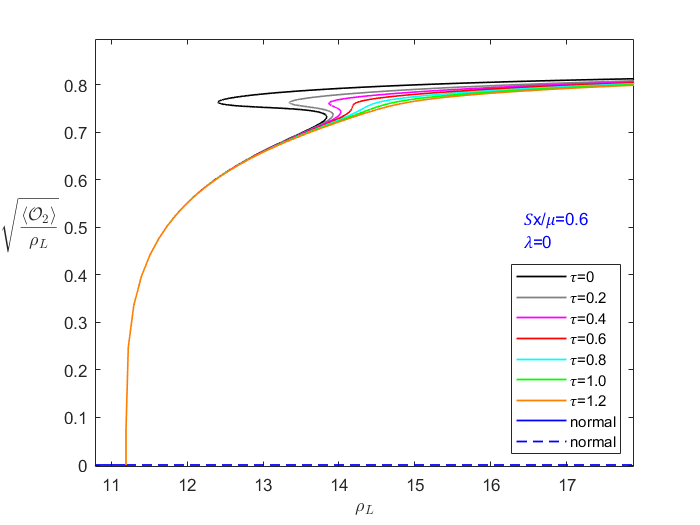}
\caption{The condensate curves in the grand canonical ensemble (\textbf{Left}) and in the canonical ensemble (\textbf{Right}) with $\lambda=0$ and different values of $\tau$.}\label{2_tau_MuRho}
\end{figure}

With the same fixed value of the superfluid velocity $S_x=0.6$, we also plot the condensate curves with various values of $\tau$ and to show the influence of the sixth power term in Figure~\ref{2_tau_MuRho}. The left panel shows the case in the grand canonical ensemble, where we observe that the condensate curve also moves rightwards with increasing values of $\tau$, which is similar to the influence of $\lambda$. However, an important difference is that even a very large value of $\tau$ would not change the superfluid phase transition from 2nd order to 1st order. The right panel of the case in the canonical ensemble shows that the sixth power term has a similar influence on the condensate curve as the fourth power term, which would remove the 1st order phase transition between the superfluid phases with higher and lower condensates.

It should be noticed that $\lambda$ and $\tau$ are the coefficients of the nonlinear terms of the scalar field $\psi$, therefore their various values do not change the critical point of the superfluid phase transition from the normal phase, where the condensate of the scalar is infinitesimal. Therefore these two terms would not change the boundary of the superfluid phase when the superfluid phase transition is still 2nd order.

Both the fourth and sixth power terms are powerful in controlling the superfluid phase transition with finite superfluid velocity, and the qualitative lawers are similar as in the case with vanishing superfluid velocity~\cite{Zhao:2022jvs,Zhao:2024jhs}. With these two powerful terms with coefficients $\lambda$ and $\tau$, let us consider how to improve the phase structure with linear instability from negative susceptibility. In order to find the region with negative susceptibility, we need the condensate grows to the different directions in the grand canonical ensemble and in the canonical ensemble, which indicates preserving the 1st order phase transition in the grand canonical ensemble. However, we would like to remove the ``cave of wind'' behavior to simplify the final phase diagram in the canonical ensemble. Comparing the influences of $\lambda$ shown in Figure~\ref{2_lam_MuRho} and that of $\tau$ in Figure~\ref{2_tau_MuRho}, we see a convenient choice of taking a large value of $\tau$ to remove the ``cave of wind'' behavior in the canonical ensemble while preserving the 1st order phase transition in the grand canonical ensemble. At the same time, the value of $\lambda$ is able to be set to 0 in a simplest set up.

\begin{figure}
\includegraphics[width=0.45\columnwidth]{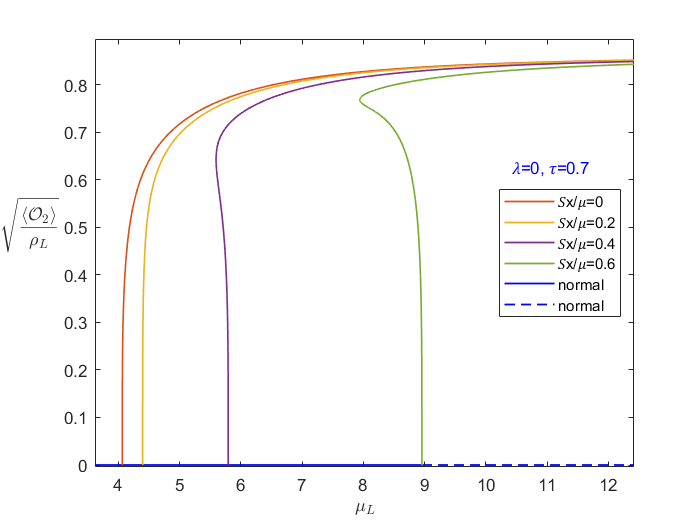} 
\includegraphics[width=0.45\columnwidth]{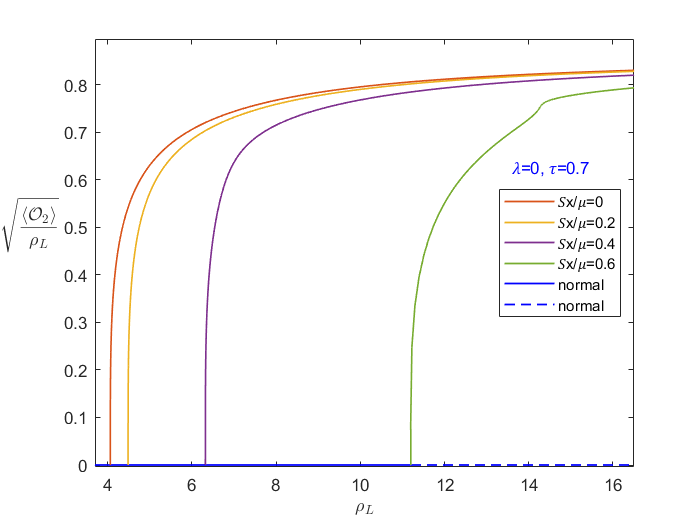}
\caption{The condensate curves in the grand canonical ensemble (\textbf{Left}) and in the canonical ensemble (\textbf{Right}) with $\lambda=0$, $\tau=0.7$ and different values of $S_{x}/\mu$.
}\label{3_O_Mu_Rho}
\end{figure}

Based on the above analysis, we take $\lambda=0$ and $\tau=0.7$ to get an improved phase diagram. We plot the condensate curves in the grand canonical ensemble as well as the canonical ensemble in Figure~\ref{3_O_Mu_Rho}. From the left panel for the condensate curves in the grand canonical ensemble, we see that the superfluid phase transition still becomes 1st order at a large value of the superfluid velocity. From the right panel of the case in the canonical ensemble, we see that there is no longer any first order phase transition between the superfluid phases with higher and lower condensate with $S_x<0.6\mu$.

\begin{figure}
\includegraphics[width=0.45\columnwidth]{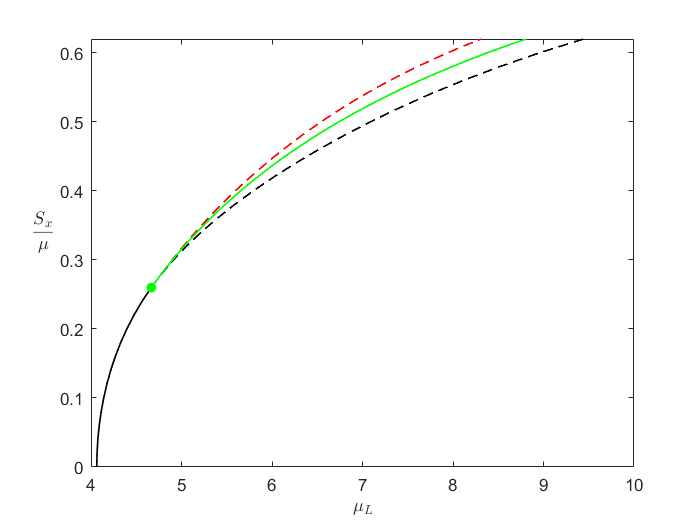}
\includegraphics[width=0.45\columnwidth]{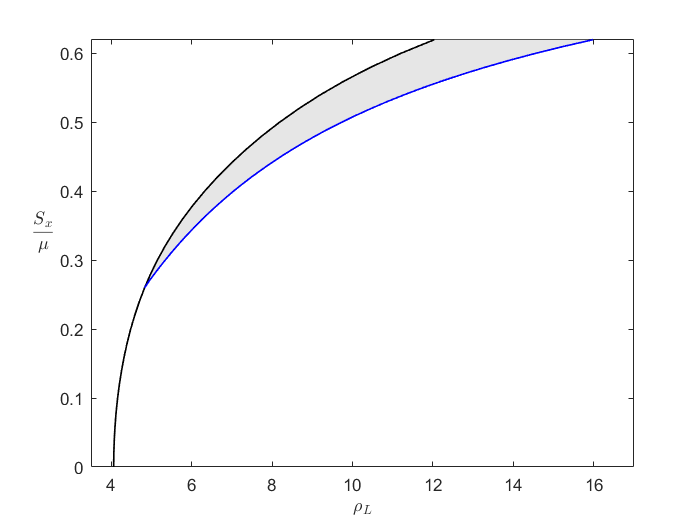}
\caption{The phase diagrams with $\lambda=0$ and $\tau=0.7$. 
\textbf{Left:}
The $S_x-\mu_L$ phase diagram in the grand canonical ensemble.
The solid black line indicates the critical points for the 2nd order superfluid phase transitions. The solid green line indicates the phase transition points for the first order superfluid phase transitions, while the dashed black and dashed red lines indicate the quasi-critical points and the turning points, respectively.
\textbf{Right:}The $S_x-\rho_L$ phase diagram in the canonical ensemble.
The solid black line indicates the position of the critical points of the 2nd order superfluid phase transitions. The solid blue line is mapped from the dashed red line in the left panel which indicates the turning points of the condensate curves in the grand canonical ensemble.}
 \label{3_Ph_Rho}
\end{figure}

With the information of the condensate curves at different values of the superfluid velocity, we are able to construct the $S_x-\mu$ phase diagram in the grand canonical ensemble and the $S_x-\rho$ phase diagram in the canonical ensemble. We plot the two phase diagrams with $\lambda=0$ and $\tau=0.7$ in Figure~\ref{3_Ph_Rho}. The left panel of the $S_x-\mu$ phase diagram show similar structure as the case with vanishing $\lambda$ and $\tau$ because that increasing the value of $\tau$ would not change the superfluid phase transition in the grand canonical ensemble from 1st order to 2nd order. However, a large value of $\tau$ would remove the ``cave of wind'' behavior, resulting in a more elegant phase diagram without the tip region with overlapping solutions, as seen in the middle and right panels in Figure~\ref{1_ph_MuRho}. The dashed red line indicating the turning point of the condensate curve in the grand canonical ensemble in the left panel of Figure~\ref{3_Ph_Rho} is therefore mapped into the solid blue line in the right panel. And finally the gray region between the solid black line and the solid blue line in the right panel represents the domain with negative susceptibility, which is useful for realizing possible turbulence or vortex structures from an initial state with a large uniform superfluid velocity.
\section{Conclusion and outlooks}\label{sect:conclusion}
In this paper, we compared the condensate curves in the grand canonical ensemble and in the canonical ensemble, from which we obtained the linear thermodynamic stability of the holographic superfluid model with finite superfluid velocity in the probe limit. We also considered the fourth and sixth power nonlinear potential terms to improve the final phase diagram in the canonical ensemble to show the region of negative susceptibility more clearly.

Without the nonlinear terms, it is found that the condensate curve shows a ``cave of wind'' behavior with large values of superfluid velocity, which results in a complex region with overlapping superfluid solutions in the $S_x-\rho$ phase diagram. The region with linear instability, resulting from negative susceptibility also involves these overlapping solutions and becomes complicated. To simplify both the phase diagram and the region with negative susceptibility, we applied the powerful fourth and sixth power terms to regulate the condensate curves.

We systematically studied the influence of the two parameters $\lambda$ and $\tau$ of the two nonlinear terms on the condensate curves in both the grand canonical ensemble and in the canonical ensemble. Our study shows that these two terms have qualitatively the same effect on the condensate curves as in the case with zero value of superfluid velocity $S_x=0$ studied in Refs.~\cite{Zhao:2022jvs,Zhao:2024jhs}. With increasing values of both the two coefficients $\lambda$ and $\tau$, the condensate curves move rightwards with the critical point unchanged. The term with the coefficient $\lambda$ is able to change the order of the superfluid phase transition, while the influences of the term with coefficient $\tau$ are more focused on the region with larger condensate.

With these two powerful terms, we are able to simplify the final $S_x-\rho$ phase diagram as well as the region with negative susceptibility. As a concrete example, we set $\lambda=0$ and $\tau=0.7$ to show an improved final phase diagram in which the region with the linear instability from the negative susceptibility is greatly simplified.

Our results reveal the linear instability which is believed to be related to the inhomogeneous perturbations~\cite{Zhao:2022jvs,Zhao:2023ffs} in the canonical ensemble with a fixed total charge for the superfluid solution with finite superfluid velocity. It is therefore important to calculate the Quasi-Normal Modes to locate the dynamical modes more precisely. Finally it would be very interesting to apply the full time dependent evolution from a homogeneous initial state in the region with negative susceptibility to see how this initial state with homogeneous superfluid velocity is deformed by the inhomogeneous modes, where formation of vortexes and turbulence are possible to be observed.
\section*{Acknowledgements}
ZYN would like to thank Hua-Bi Zeng and Chuan-Yin Xia for valuable discussions.
This work is partially supported by the National Natural Science Foundation of China (Grant Nos.11965013).
ZYN is partially supported by Yunnan High-level Talent Training Support Plan Young $\&$ Elite Talents Project (Grant No. YNWR-QNBJ-2018-181).

\bibliographystyle{apsrev4-1}
\bibliography{reference}
\end{document}